\documentclass[conference]{IEEEtran}
\IEEEoverridecommandlockouts

\usepackage{cite}
\usepackage{amsmath,amssymb,amsfonts}
\usepackage{bbm}
\usepackage{algorithm}
\usepackage{algorithmicx}
\usepackage[noend]{algpseudocode}
\usepackage{tabularx}
\usepackage{graphicx}
\usepackage{textcomp}
\usepackage{xcolor}
\usepackage{enumitem}
\usepackage{booktabs}
\usepackage{url}

\newtheorem{definition}{Definition}[section]

\def\BibTeX{{\rm B\kern-.05em{\sc i\kern-.025em b}\kern-.08em
    T\kern-.1667em\lower.7ex\hbox{E}\kern-.125emX}}
\begin{document}

\title{Accelerating High-Dimensional Nearest Neighbor Search with Dynamic Query Preference}


\author{\IEEEauthorblockN{Yifan Zhu\IEEEauthorrefmark{1},
Ruijie Zhao\IEEEauthorrefmark{1},
Zhonggen Li\IEEEauthorrefmark{1},
Baihua Zheng\IEEEauthorrefmark{2},
Junyi Qiu\IEEEauthorrefmark{1},
Zhikun Zhang\IEEEauthorrefmark{1},
 and
Congcong Ge\IEEEauthorrefmark{1}}
\IEEEauthorblockA{\IEEEauthorrefmark{1}Zhejiang University, Hangzhou, China\\
Email: \{xtf\_z, ruijie.zhao, zgli, junyiqiu, zhikun, gcc\}@zju.edu.cn}
\IEEEauthorblockA{\IEEEauthorrefmark{2}Singapore Management University, Singapore\\
Email: bhzheng@smu.edu.sg}

}

\maketitle

\begin{abstract}
Approximate Nearest Neighbor Search (ANNS) has emerged as an essential operation in modern database and AI systems. While graph-based methods like NSG demonstrate state-of-the-art ANNS performance, they typically ignore that query distributions are often skewed. In real-world scenarios, user preferences and time-varying access patterns lead to non-uniform workloads, where specific data regions are retrieved significantly more frequently than others. Meanwhile, these patterns evolve over time, making pre-built indexes outdated and thus inefficient for future query workloads. Motivated by this, we propose \textbf{DQF}, a novel \textbf{\underline{D}}ual-Index \textbf{\underline{Q}}uery \textbf{\underline{F}}ramework for dynamic query preference. This dual-index structure comprises a Hot Index containing frequently accessed nodes and a Full Index covering the entire dataset, so that hot queries can be answered faster within the compact Hot Index while cold queries still obtain complete results from the Full Index. Furthermore, we propose a three-phase competitive search in which both layers share a single priority queue. A lightweight decision tree detects when the top-$k$ results have stabilized and triggers per-query early termination. To address temporal shifts in query patterns, we design an adaptive update mechanism that periodically promotes new high-frequency nodes to the Hot Index while demoting outdated ones. Experiments on five real-world datasets demonstrate that DQF achieves a 2.2--6.9$\times$ speedup over the strongest baseline on each million-scale dataset at 95\% recall. Moreover, it scales to 100M vectors with consistent performance gains, successfully adapting to distribution shifts without requiring Full Index reconstruction.

\end{abstract}

\begin{IEEEkeywords}
Dual-Index, Nearest Neighbor Search, Query Preference
\end{IEEEkeywords}

\section{Introduction}
Nearest Neighbor Search (NNS) in high-dimensional spaces supports a wide range of applications, including information retrieval \cite{infor-retrivalEx1, infor-retrivalEx2, infor-retrivalEx3}, recommendation systems \cite{RecommendEx1, RecommendEx2, RecommendEx3}, and retrieval-augmented generation \cite{RAGEX1, RAGEX2, RAGEX3}. Due to the curse of dimensionality \cite{CurseofDemEx1, CurseofDemEx2}, exact NNS is computationally expensive for large-scale datasets. To address this issue, numerous approximate NNS methods have been proposed, including hash-based \cite{QALSH, iDEC, VHP}, quantization-based \cite{PQ, VAQ, DeltaPQ}, tree-based \cite{VPTree, KDTree, RTree}, and graph-based methods \cite{Graph-servey, DEG, taoMNG, InnerProduct, InnerProduct2}. Among these, graph-based methods~\cite{NSW, HNSW, NSG, NSSG, Diskann} achieve the best balance between search speed and recall~\cite{wang2024starling}.

However, existing graph-based methods share a common premise: the graph topology is shaped purely by \textit{geometric} proximity, and every data point is assumed to be queried with equal probability. In practice, queries are far from uniformly distributed. Instead, they follow Zipf's law~\cite{Zipflaw}, where an item's frequency is inversely proportional to its rank. This skew is widespread in information retrieval and recommendation systems that increasingly rely on embedding-based retrieval~\cite{EBRFacebook, YouTubeDNN, EBRTaobao}. For instance, a small fraction of popular videos account for the majority of views on YouTube~\cite{ZipfYoutube}, while query frequencies obey a power-law distribution in web search engines~\cite{PowerLawIR}. Moreover, item popularity is heavily long-tailed, with a small fraction of items attracting the overwhelming majority of user interactions for e-commerce recommendations~\cite{ParkLongTail}.

\begin{figure}
  \centering
  \includegraphics[width=\linewidth]{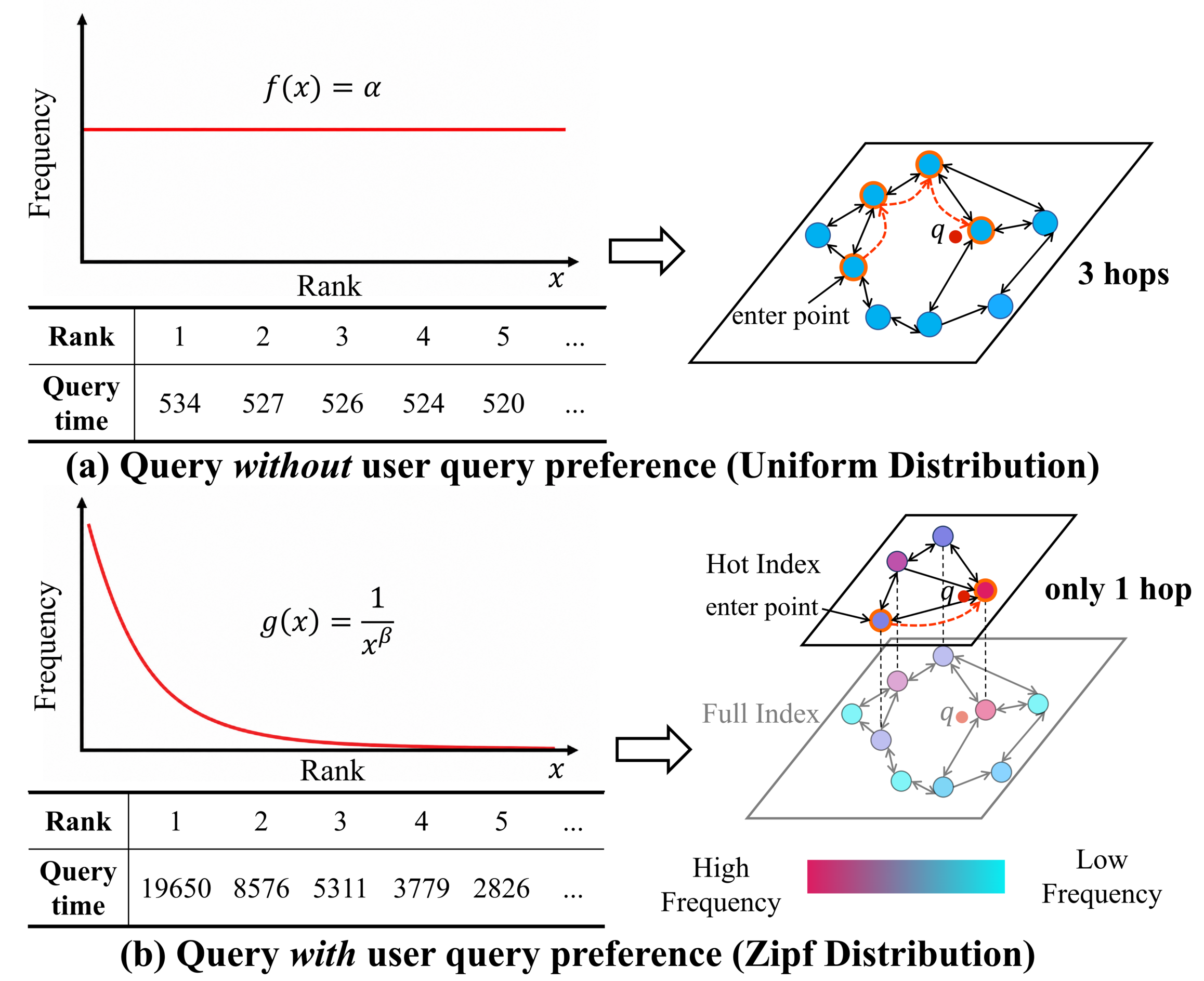}
  \caption{Query Distribution Patterns: Uniform vs. Zipf. The ``1 hop vs.\ 3 hops'' contrast is illustrative, depicting how placing frequently accessed vectors near the entry point shortens the \emph{traversal} to a popular target.}
  \label{fig:intro}
\end{figure}

The popularity skew transfers to the nearest-neighbor workload. This motivates an optimization in graph index construction, where the index layout should be shaped by workload frequency, not only by geometric proximity. As Figure~\ref{fig:intro} illustrates, placing frequently accessed vectors closer to the entry point reduces traversal from 3 hops to 1 (Figure~\ref{fig:intro}a vs.\ b). However, building a graph index that exploits this frequency skew poses three challenges.

\vspace{0.1cm}
{\textbf{\textit{Challenge I: How to cache hot data in high-dimensional space?}}
Traditional caches rely on exact key matching to reuse previous query results, but semantically identical queries produce different embedding vectors, so query-level caching achieves near-zero hit rates. Since the frequency skew resides in \textit{target objects} rather than query vectors, the cache must be organized around object-level frequency.
To address this, we propose a dual-layer index structure. The upper layer, termed the \textit{Hot Index}, stores high-frequency target nodes, while the lower layer, called the \textit{Full Index}, maintains the complete dataset to ensure coverage. During construction of the Hot Index, we further apply a lightweight \textit{frequency-aware bias} during neighbor selection to favor high-frequency nodes.
}

\textbf{\textit{Challenge II: How to adapt to shifting query patterns?}}
Query distributions evolve continuously, so the Hot Index built for yesterday's workload may no longer match today's. Since full graph reconstruction is prohibitively expensive~\cite{Diskann}, adaptation should ideally be confined to the Hot Index. Yet even this is non-trivial, as three closely related questions arise.
(i) \textit{When} to update, since triggering too late lets hit rates degrade while triggering too frequently wastes compute on short-lived fluctuations.
(ii) \textit{What} to update, since naive swapping of borderline nodes causes frequent topology changes and impairs search quality.
(iii) \textit{How} to update, since removing a node damages the connectivity of its neighbors and requires careful repair of the surrounding topology.
To address these, DQF employs three mechanisms. First, a sliding-window hit-rate detector identifies distribution drift online. Second, an \textit{asymmetric buffer} applies a stricter threshold for insertion than for deletion to prevent frequent boundary changes. Third, a \textit{layered repair} scheme handles two cases separately: lightly damaged nodes are repaired through lightweight edge borrowing, while severely damaged neighborhoods are reconstructed with full angle-based pruning.

\vspace{0.1cm}
\textbf{\textit{Challenge III: How to organize the search across two indices?}}
For dual-layer indexes, the prevailing strategy first searches the Hot Index, then continues on the Full Index. However, for high-frequency queries, the Hot Index alone suffices to return high-quality top-$k$ candidates, making the Full Index continuation unnecessary. For low-frequency queries, the Hot Index contains few relevant nodes, so traversing it largely wastes time, and the Full Index must explore extensively. 
To address this, we introduce a three-phase \textit{competitive search}. It first obtains a high-quality seed from the Hot Index and locates a corresponding entry in the Full Index, then merges candidates from both layers into a single distance-ordered queue for source-aware competitive expansion. This design exploits the cache locality of the compact Hot Index for frequent queries while preserving Full Index coverage for infrequent ones. On top of this, a lightweight \textit{decision tree} inspects runtime search-state features to predict per-query stopping points, terminating early for frequent queries while continuing exploration for rare ones.

In summary, we make the following key contributions:
\begin{itemize}[leftmargin=*]
\item{} \textit{Dual-Index Query Framework.} We propose \textbf{DQF}, a framework that optimizes graph-based ANNS for frequency-based query patterns, allocating more search effort to data points that are actually retrieved most often.
\vspace{0.03in}
\item{} \textit{Semantic Caching via Hot Index.} We design a two-layered index structure that captures frequency-based hotspots. The upper layer stores high-frequency nodes as a semantic cache, with a frequency-weighted construction that shortens hit paths without distorting the topology. The lower layer maintains the complete dataset for full coverage.
\vspace{0.03in}
\item{} \textit{Adaptive Update under Query Shift.} We propose a hit-rate-triggered update scheme combining an asymmetric buffer and layered repair, which keeps the Hot Index aligned with drifting workloads without rebuilding the Full Index.
\vspace{0.03in}
\item{} \textit{Competitive Search Strategy.} We devise a source-aware competitive search expansion strategy and an early termination mechanism, which jointly schedule high-quality candidates and adaptively reduce the search depth.
\vspace{0.03in}
\item{} \textit{Extensive Experiments.} Experiments on five real-world datasets show that DQF achieves 2.2$\times$ to 6.9$\times$ speedup over the strongest baseline at the million scale and 1.3$\times$--1.7$\times$ at the hundred-million scale, while maintaining 95\% recall. Notably, DQF requires no Full Index reconstruction under changing query distributions.
\end{itemize}

The rest of this paper is organized as follows. We review previous work in Section~\ref{sec:2-rel} and present the problem statement in Section~\ref{sec:3-back}. Subsequently, we introduce our newly proposed dual-index query framework, DQF, in Section ~\ref{sec:4-index}. Then we describe the competitive search with early stopping in Section~\ref{sec:5-search}, and address the adaptive hot index maintenance in Section~\ref{sec:6-maint}. Finally, we report comprehensive experimental studies in Section~\ref{sec:7-exp} and conclude the paper in Section~\ref{sec:8-concl}.

\section{Related Work}
\label{sec:2-rel}

In this section, we review the existing work on nearest neighbor search and optimizations for dynamic workloads. 

\subsection{Non-Graph-Based Methods}
Non-graph-based ANNS methods fall into three families, i.e., hash-based methods, quantization-based methods, and tree-based methods. Hash-based methods such as LSH~\cite{LSH} map vectors to low-dimensional hash codes for fast bitwise lookup. Quantization-based methods, including PQ~\cite{PQ} and OPQ~\cite{OPQ}, compress vectors into compact codes to reduce storage and computation. Tree-based methods such as KDTree~\cite{KDTree} and VPTree~\cite{VPTree} partition the space hierarchically. All three families degrade in high-dimensional spaces: hash collisions and quantization errors degrade result quality, while tree partitions become ineffective at narrowing the search space~\cite{CurseofDemEx1}.

\subsection{Graph-Based Methods}
In contrast, graph-based methods construct a proximity graph over data points and answer queries via greedy traversal. For instance, NSW~\cite{NSW} builds a navigable small-world graph, HNSW~\cite{HNSW} adds a hierarchical skip-list structure for logarithmic search, NSG~\cite{NSG} prunes edges via Monotonic Relative Neighborhood Graph theory, while NSSG~\cite{NSSG} further balances out-degree distribution with its satellite system graph pruning. These methods achieve the best speed--recall trade-offs on large-scale, high-dimensional benchmarks~\cite{wang2025accelerating}, yet all assume uniform query distributions, limiting effectiveness under skewed or shifting workloads.

\subsection{Workload-Aware Indexing}
\label{sec:2-rel-dynamic}

Real-world query distributions are rarely uniform, typically following a Zipfian pattern where a small fraction of data attracts the majority of traffic~\cite{ZipfGoogle, ZipfYoutube, ZipfRAG}. Therefore, several methods adapt index structures by exploiting temporal locality. For instance, PANNS~\cite{PANNS} reorganizes the graph around recently \emph{inserted} objects, modeling hotness as a Gaussian distribution centered on insertion time. In order to track shifting data distributions, DeDrift~\cite{Dedrift} incrementally updates IVF coarse-quantizer centroids without full index rebuilds. However, existing approaches presume future queries are dominated by newly ingested or recently accessed data. They are ineffective in scenarios where long-resident items stay consistently popular independent of insertion time. Such drawbacks underscore the necessity of developing dedicated techniques for content-driven query preferences adhering to specific distributions, such as Zipfian distribution.

\section{Background}
\label{sec:3-back}

In this section, we formalize the approximate nearest neighbor search problem and then characterize the Zipfian access patterns that motivate our dual-index design.

\subsection{Problem Formulation}
We first define exact nearest neighbor search (NNS), which retrieves the $k$ closest data points to a query:

\begin{definition}[Nearest Neighbor Search]
Given a dataset \( \mathcal{D} = \{ \mathbf{x}_1, \mathbf{x}_2, \dots, \mathbf{x}_n \} \) where \( \mathbf{x}_i \in \mathbb{R}^d \), a query point \( \mathbf{q} \in \mathbb{R}^d \), and a distance metric \( \text{dist}(\cdot, \cdot) \), the nearest neighbor search seeks a subset \( \mathcal{N}_k(\mathbf{q}) \) from \( \mathcal{D} \) that are closest to \( \mathbf{q} \):
\begin{equation} 
\small
\mathcal{N}_k(\mathbf{q}) = \arg\min_{\mathcal{S} \subseteq \mathcal{D}, |\mathcal{S}|=k} \sum\nolimits_{\mathbf{x} \in \mathcal{S}} \text{dist}(\mathbf{q}, \mathbf{x}).
\end{equation}
\end{definition} 

Exact NNS is computationally prohibitive in high dimensions, so we focus on approximate $k$-nearest neighbor search, which is defined as follows:

\begin{definition}[$\epsilon$-Nearest Neighbor Search]
Given a dataset \( \mathcal{D} = \{ \mathbf{x}_1, \mathbf{x}_2, \dots, \mathbf{x}_n \} \) where \( \mathbf{x}_i \in \mathbb{R}^d \), a query point \( \mathbf{q} \in \mathbb{R}^d \), a distance metric \( \text{dist}(\cdot, \cdot) \), and an approximation factor \( \epsilon > 0 \), let \( \mathbf{x}^*_1, \dots, \mathbf{x}^*_k \) denote the exact $k$ nearest neighbors of \(\mathbf q\) sorted by ascending distance. The $\epsilon$-nearest neighbor search returns an ordered set \( \mathcal{A}_k(\mathbf{q}) = \{\mathbf a_1, \dots, \mathbf a_k\} \subseteq \mathcal{D} \), also sorted by ascending distance, such that each returned point is within a \((1+\epsilon)\) factor of the corresponding true neighbor:
\begin{equation}
\small
\text{dist}(\mathbf{q}, \mathbf{a}_i) \;\leq\; (1 + \epsilon)\, \text{dist}(\mathbf{q}, \mathbf{x}^*_i), \qquad i = 1, \dots, k.
\end{equation}
\end{definition}

\begin{figure*}
  \centering
  \includegraphics[width=\linewidth]{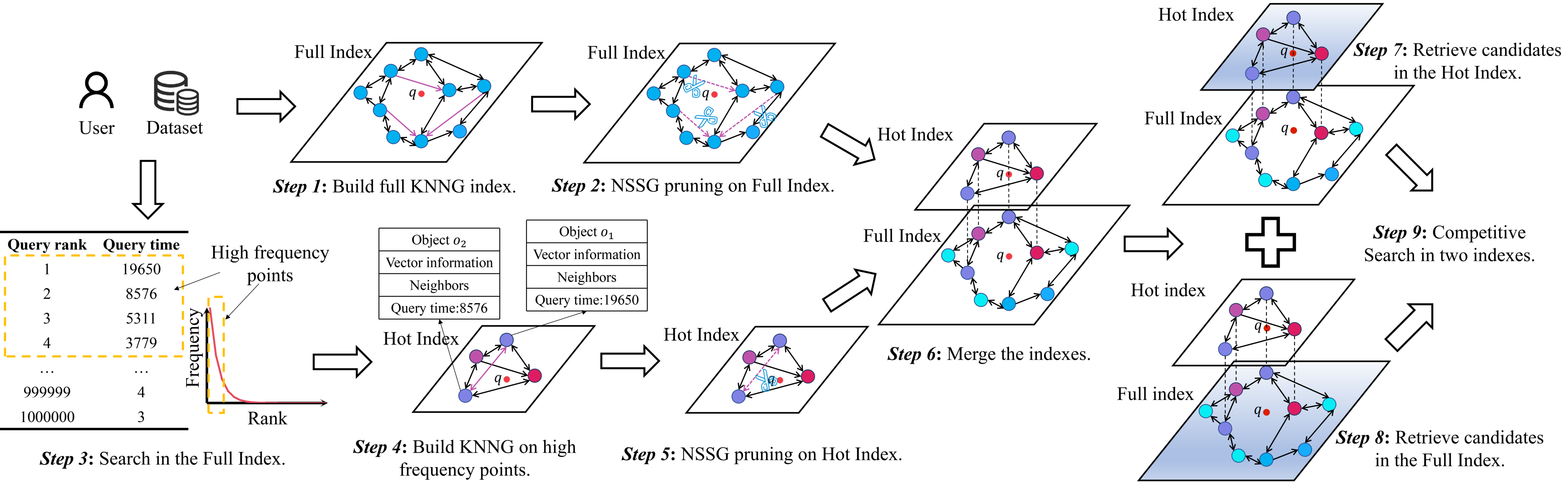}
  \vspace{-0.2in}
  \caption{Overview of the Dual-Index Query Framework}
  \label{fig:struct}
  \vspace{-0.1in}
\end{figure*}

\subsection{Zipfian Access Patterns}
Zipf's law states that an item's frequency is inversely proportional to its rank~\cite{Zipflaw}:
\begin{equation}
\small
f(r) \sim r^{-\alpha}
\end{equation}
where \(r\) is the frequency rank, \(\alpha\) denote the skewness exponent, and \(f(r)\) denote the occurrence frequency. We use $\alpha$ consistently for the exponent throughout the paper.

Beyond natural-language corpora, Zipf's law also governs data-access patterns in real-world systems. Clauset et al.~\cite{Zipf24Dataset} confirmed it across 24 datasets, Gill et al.~\cite{ZipfYoutube} showed that user attention on YouTube concentrates on a small fraction of videos, and web-page visits and search clicks follow the same power-law~\cite{ZipfGoogle, ZipfWeb}. These findings indicate that the access frequency of target objects follows Zipf's law.

This creates a mismatch with the uniform-access assumption of most ANNS indexes, i.e., a small fraction of data receives the majority of queries, while the vast tail is rarely accessed.

\section{Dual-layer Index Construction}
\label{sec:4-index}
In this section, we present an overview of our dual-layer index DQF, describe the hot data identification, introduce the dual-layer construction, and provide a theoretical analysis of space consumption and time complexity.

\subsection{Overview}
As a small fraction of target objects attracts the vast majority of retrievals under Zipf-distributed queries, the proposed index DQF responds to this skew of real workloads by constructing a compact \emph{Hot Index} over the most frequently retrieved nodes, and a \emph{Full Index} over the entire dataset. Both layers are instances of the Navigating Satellite System Graph (NSSG)~\cite{NSSG}. Specifically, the Full Index $G_{\text{full}}$ is a standard NSSG over the entire dataset $D$ ($|D| = N$), guaranteeing coverage for cold queries. The Hot Index $G_{\text{hot}}$ is a much smaller NSSG over a hot subset $D_{\text{hot}} \subset D$ of size $N_h \ll N$ (typically $N_h = 0.005 \cdot N$), serving as a cheap entry point for frequent queries. Since graph search runs in expected $O(\log n)$ time, restricting initial exploration to a subgraph roughly two orders of magnitude smaller eliminates most unproductive hops. Figure~\ref{fig:struct} illustrates the overall framework.

\subsection{Hot Identification}
\label{sec:hot-id}
To construct the Hot Index, DQF must first identify which objects belong to the hot partition. Traditional caches match on exact keys, but semantically identical queries produce distinct embedding vectors, making query-level caching impractical. Therefore, DQF defines hotness at the \emph{target-object} level: during online serving on $G_{\text{full}}$, the system records how often each node appears in query results. For a historical workload $Q_{\text{his}}$, the accumulated frequency of node $v$ is:
\begin{equation}
f(v) \;=\; \sum_{q \in Q_{\text{his}}} \mathbbm{1}\!\left[v \in \mathrm{TopK}(q)\right].
\end{equation}

The $N_h$ nodes with the largest $f(v)$ are selected as $D_{\text{hot}}$ and used to construct the Hot Index $G_{\text{hot}}$.

\subsection{Frequency-Weighted Hot Subgraph}
\label{sec:freq-prune}
Standard NSSG construction sorts candidate neighbors by distance alone and applies an angle-based diversity test, treating every node identically. However, in the hot index $G_{\text{hot}}$, a small group of highly popular nodes dominates the search targets. Owing to their higher probability of being selected as graph neighbors, greedy search reaches these frequent targets in fewer steps.

Thus, we incorporate retrieval frequency into the neighbor-selection process as a distance-scaling weight. Due to raw frequencies $f(v)$ varying by several orders of magnitude across nodes, we first compress them with a logarithm and normalize to $[0,1]$:
\begin{equation}
\begin{aligned}
F(v) &\;=\; \frac{\ln f(v) - \ln f_{\min}}{\ln f_{\max} - \ln f_{\min}}, \\[2pt]
w(v) &\;=\; 1 - \beta\, F(v)^{p},
\end{aligned}
\label{eq:weight}
\end{equation}
with $\beta = 1 - w_{\min}$. Defaults are $w_{\min} = 0.8$ and $p = 1$, capping the discount at $20\%$. The weight enters the pool-ordering comparator:
\begin{equation}
p_i \prec p_j \;\iff\; d(v, p_i)\, w(p_i) \;<\; d(v, p_j)\, w(p_j).
\label{eq:cmp}
\end{equation}

Intuitively, a node with higher query frequency incurs lower weighting coefficients $w$, making its weighted distance $d \cdot w$ shorter and giving it priority in the candidate pool. Since NSSG pruning examines candidates sequentially, an earlier candidate faces fewer already-selected neighbors and is more likely to pass the angle test. Therefore, popular nodes gain more opportunities to be selected into $N(v)$.
\begin{algorithm}[tbp]
\caption{Dual-layer Index Construction}
\label{alg:construction}
\begin{algorithmic}[1]
\Require dataset $D$, frequency counters $f(\cdot)$, hot size $N_h$, max degree $M$, angle threshold $\theta$.
\Ensure dual-layer index $(G_{\text{full}}, G_{\text{hot}})$.
\State $G_{\text{full}} \gets \textsc{NSSG}(D, M, \theta)$
\State $D_{\text{hot}} \gets$ top-$N_h$ nodes by $f(\cdot)$
\State compute $w(\cdot)$ from $f(\cdot)$ \Comment{Eq.~\eqref{eq:weight}}
\State $G_{\text{hot}} \gets \textsc{WeightedNSSG}(D_{\text{hot}}, M, \theta, w)$ \Comment{Eq.~\eqref{eq:cmp}}
\State \Return $(G_{\text{full}}, G_{\text{hot}})$
\end{algorithmic}
\end{algorithm}

Algorithm~\ref{alg:construction} summarizes the construction pipeline. First, a standard NSSG is built over the entire dataset $D$ to obtain the Full Index $G_{\text{full}}$ (line~1). The frequency counters $f(v)$, accumulated during online serving, record how often each node appears in query results. The $N_h$ nodes with the highest frequencies form the hot subset $D_{\text{hot}}$ (line~2). The frequency weights $w(\cdot)$ are then derived via Eq.~\eqref{eq:weight} (line~3), and a frequency-weighted NSSG is constructed over $D_{\text{hot}}$ to produce the Hot Index $G_{\text{hot}}$ (line~4). The $\textsc{WeightedNSSG}$ routine follows standard NSSG construction~\cite{NSSG}, replacing raw-distance ordering with the frequency-weighted comparator of Eq.~\eqref{eq:cmp}.

\subsection{Complexity Analysis}
\label{sec:IdxComplex}

In this subsection, we analyze the space and time cost of constructing both layers.

\noindent\textbf{Space.}
Each layer stores a $d$-dimensional vector and a neighbor list of up to $M$ entries per node, for a total of $O((N + N_h)(d + M))$.

\noindent\textbf{Time.}
Building $G_{\text{full}}$ follows standard NSSG construction~\cite{NSSG} and dominates the overall cost. Building $G_{\text{hot}}$ runs the same NSSG pipeline on $D_{\text{hot}}$ at scale $N_h \ll N$, so its cost is negligible relative to $G_{\text{full}}$. The frequency counters are collected during online serving at no extra construction cost. The weighting only adds one multiplication per candidate comparison, leaving the asymptotic complexity unchanged. Overall, the construction overhead of DQF beyond a single NSSG build is marginal.

\section{Competitive Search with Early Stopping}
\label{sec:5-search}
Standard graph-based ANNS uses beam search~\cite{NSG, NSSG, HNSW} that greedily expands a priority queue of $L$ candidates until no unvisited candidate remains. Consequently, every query exhausts the full budget regardless of difficulty. A dual-layer index further introduces a scheduling challenge, where frequent queries can often be resolved within the compact hot graph alone, while rare queries must reach into the full graph. Searching both layers indiscriminately for every query wastes effort in either direction. This raises two coupled questions: how to coordinate the exploration of two graphs, and when to stop for each individual query.

To address these problems, we propose a three-phase competitive search on the dual-layer index, and develop a lightweight decision tree that stops each query individually once its result has converged.

\subsection{Three-Phase Competitive Search}
\label{sec:competitive-search}

The straightforward two-pass strategy exhaustively traverses the Hot Index before invoking the Full Index, which incurs redundant computation. When the Hot Index already contains the answer, the Full-Index pass is redundant. Otherwise, the Full Index search starts without any useful seed and discards the progress the Hot Index made. To avoid both pitfalls, DQF lets the two layers \emph{cooperate} within a single expanding queue through the following three phases, which gradually widen the exploration from the hot graph to the full graph.

\noindent\textbf{Phase 1 (Hot Local Search).}
Phase~1 exploits the compact hot graph to quickly produce high-quality seed candidates for the subsequent full-graph exploration. The queue is seeded with the entry point of $G_{\text{hot}}$ and several random hot nodes. Greedy expansion proceeds inside $G_{\text{hot}}$ only, and terminates when the next candidate to expand is farther from $q$ than the current Top-$K'$, where $K'$ is a small stabilization depth (default $K'=1$). As $G_{\text{hot}}$ is roughly two orders of magnitude smaller than $G_{\text{full}}$ and $K'$ is kept small, this phase produces a high-quality seed with very few distance computations.

\noindent\textbf{Phase 2 (Full Stabilization).} In this phase, the top-$K'$ entries are expanded against $G_{\text{full}}$ until they become stable in the full graph. This ensures that the leading candidates reflect both layers before competitive expansion starts. 

\noindent\textbf{Phase 3 (Source-Aware Competitive Expansion).}
After stabilization, candidates from both layers reside in a single priority queue ranked by raw distance to $q$, so the two graphs now \emph{compete} on equal footing. The queue is traversed in distance order. For a candidate $v$ at position $k$, the routing function $\mathrm{expand}(v)$ decides which graph to retrieve $v$'s neighbors from:
\begin{equation}
\mathrm{expand}(v) = \begin{cases} G_{\text{hot}}, & v \in D_{\text{hot}}\ \text{and}\ k < \tau_{\text{switch}}, \\ G_{\text{full}}, & \text{otherwise}, \end{cases}
\label{eq:route}
\end{equation}
with $\tau_{\text{switch}} = K + (L-K)/c$, where $c$ defaults to $3$. Beyond $\tau_{\text{switch}}$, the limited coverage of the hot subgraph offers diminishing returns and expansion falls back to $G_{\text{full}}$.

\subsection{Decision Tree-Based Early Stopping}
\label{sec:dt}

The three-phase design resolves how to schedule two layers, but every query still expands the same fixed number of candidates regardless of difficulty. Prior learned-stopping methods, such as LAET~\cite{LAET} and AdaNNS~\cite{AdaNNS} address per-query budgets but keep the index layout unchanged, so their stopping signals cannot leverage the dual-layer structure, such as hot-graph proximity. They also rely on pre-computed ground truth for training, which is expensive to refresh under a distribution shift. DQF addresses both limitations with a lightweight decision tree that predicts per-query convergence during Phase~3. The training details are described below.

\subsubsection{Feature Design}
\label{sec:dt-feats}

An effective stopping predictor must capture two complementary signals: how \emph{difficult} the query is inherently, and how far the search has \emph{converged} toward a stable result. The former is determined before competitive expansion begins and sets the expected workload, while the latter is revealed progressively during expansion and indicates whether that workload has been completed.

Each invocation of the tree consumes seven features (Table~\ref{tab:decision_tree_features}), which can be split into four groups. $f_1, f_2$ are fixed at the end of Phase 1 and capture query difficulty by measuring how far the query lies from the hot region. $f_3, f_4$ track current result quality through the gap between the running Top-1 and Top-$K$. $f_5, f_6$ measure search progress. $f_7$ reports the overlap between the running Top-$K$ and $D_{\text{hot}}$.

\begin{table}[tbp]
\centering
\caption{Decision tree features ($f_1, f_2$: static, $f_3$--$f_7$: dynamic).}
\label{tab:decision_tree_features}
\vspace{-0.1in}
\begin{tabularx}{0.48\textwidth}{lX}
\toprule
\textbf{Feature} & \textbf{Description} \\
\midrule
$f_1$: \texttt{hot\_dist\_1st} & Distance to nearest neighbor in $G_{\text{hot}}$ at the end of Phase 1. \\
$f_2$: \texttt{hot\_ratio\_1\_k} & $f_1$ divided by the distance to the $K$-th nearest in $G_{\text{hot}}$. \\
$f_3$: \texttt{dist\_1st} & Distance to the current Top-1 in $\mathrm{retset}$. \\
$f_4$: \texttt{dist\_ratio\_1\_k} & $f_3$ divided by the distance to Top-$K$. \\
$f_5$: \texttt{dist\_cnt} & Cumulative distance computations so far. \\
$f_6$: \texttt{update\_cnt} & Number of Top-$K$ updates so far. \\
$f_7$: \texttt{hot\_ratio\_topk} & Fraction of the current Top-$K$ that lies in $D_{\text{hot}}$. \\
\bottomrule
\end{tabularx}
\vspace{-0.12in}
\end{table}

\subsubsection{Offline Training Data Collection}
\label{sec:dt-train}

Training data is collected by replaying a random subset of $Q_{\text{his}}$ (5\,000 queries by default) on the competitive search loop. Let $\mathrm{stop\_pos}(q)$ be the distance-computation count at which $q$'s Top-$K$ last changes. During Phase~3 replay, a training sample is logged every $\mathit{check\_period}$ with label
\begin{equation}
y \;=\; \mathbbm{1}[\,c > \mathrm{stop\_pos}(q)\,].
\label{eq:label}
\end{equation}

Labels are derived from search replay rather than external ground truth, so training requires no brute-force computation, and the tree can be retrained cheaply after each drift event. However, the replay budget must exceed the search pool size $L$ used at query time. At high target recall, the actual stopping position often exceeds $L$, and capping at $L$ would mislabel late continue samples as stops. We therefore scale the budget by a recall-dependent multiplier with $k = \ln 3 / 0.09$:
\begin{equation}
L_{\text{train}} \;=\; \mu(\rho^*) \cdot L,
\label{eq:Ltrain}
\end{equation}
\begin{equation}
\mu(\rho^*) \;=\; \begin{cases} 1, & \rho^* < 0.9, \\ e^{k(\rho^* - 0.9)}, & 0.9 \le \rho^* < 0.99, \\ 3, & \rho^* \ge 0.99, \end{cases}
\label{eq:mu}
\end{equation}

\subsubsection{Three-Tier Sample Weighting}
\label{sec:dt-weight}

The collected samples are imbalanced in three ways. First, within each query, the continue ($y\!=\!0$) and stop ($y\!=\!1$) samples are unevenly split, because a query that converges early produces few continue samples and many stop samples, while a late-converging query shows the opposite pattern. Second, under a Zipf workload, easy queries that converge in a few hundred distance computations are far more common, so they contribute most of the training samples, and the tree underweights the hard queries that actually need careful stopping decisions. Third, when the target recall is high, the continue samples logged just before $\mathrm{stop\_pos}(q)$ are the rarest, yet the most important: misclassifying any of them as stop causes premature termination and directly lowers recall. DQF compensates with three multiplicative weights.

\noindent\textbf{Per-query Class Balance.} Inside a query with $n_c$ continue and $n_s$ stop samples ($n = n_c + n_s$),
\begin{equation}
w_{\text{class}}(i) \;=\; \begin{cases} n / (2 n_c), & y_i = 0, \\ n / (2 n_s), & y_i = 1. \end{cases}
\label{eq:wclass}
\end{equation}

\noindent\textbf{Query Difficulty.} Hard queries (those whose stop position exceeds the mean $\bar{s}$ over $Q_{\text{his}}$) are amplified by
\begin{equation}
w_q \;=\; \begin{cases} \mathrm{stop\_pos}(q) / \bar{s}, & \mathrm{stop\_pos}(q) \ge \bar{s}, \\ 1, & \text{otherwise}. \end{cases}
\label{eq:wq}
\end{equation}
Easy queries keep weight $1$, so the rescaling is one-sided.

\noindent\textbf{High-recall Continue Boost.} Continue samples receive an additional weight based on the target recall $\rho^*$:
\begin{equation}
w_{\text{boost}} \;=\; \begin{cases} \bigl(0.1 / (1 - \rho^*)\bigr)^{0.7}, & \rho^* \ge 0.9, \\ 1, & \rho^* < 0.9. \end{cases}
\label{eq:wboost}
\end{equation}
The final per-sample weight is the product
\begin{equation}
w(i) \;=\; w_{\text{class}}(i) \cdot w_{q(i)} \cdot \bigl[w_{\text{boost}}\ \text{if}\ y_i = 0,\ \text{else}\ 1\bigr].
\label{eq:wtot}
\end{equation}
where $w_{\text{boost}}$ applies only to continue samples ($y_i = 0$) to penalize premature stopping. We train a classification tree with weighted Gini impurity, setting the maximum depth to 10 and the minimum leaf size to 10.

\subsection{Overall Search Procedure}
\label{sec:search-procedure}

Algorithm~\ref{alg:search} integrates the three-phase competitive search with the decision tree early stopping into a unified procedure. First, Phase~1 (lines 1--3) seeds the candidate pool $\mathrm{retset}$ from the entry points of $G_{\text{hot}}$ and greedily expands within the hot graph until the expansion pointer passes position $K'\!-\!1$, at which point the top-$K'$ entries are locally stable. Next, Phase~2 (lines 4--5) re-expands the top-$K'$ entries against $G_{\text{full}}$, ensuring that the leading candidates reflect both layers before competitive expansion begins. Finally, Phase~3 (lines 6--18) performs source-aware competitive expansion on the merged queue: each unvisited candidate is routed to $G_{\text{hot}}$ or $G_{\text{full}}$ according to Eq.~\eqref{eq:route}. Every $\mathit{check\_period}$ distance computations, the decision tree $T$ is evaluated on the current feature vector. If the tree predicts convergence ($T(\mathbf{f}) = 1$), the search terminates early (line 16), allowing easy queries to exit after a few hundred distance computations while hard queries consume the full budget.

\begin{algorithm}[tbp]
\caption{Competitive Search}
\label{alg:search}
\begin{algorithmic}[1]
\Require dual-layer index $(G_{\text{full}}, G_{\text{hot}})$, query $q$, top-$K$, pool size $L$, stabilization depth $K'$, check period $\mathit{check\_period}$, decision tree $T$.
\Ensure approximate Top-$K$ of $q$.
\State $\mathrm{retset} \gets$ entry points of $G_{\text{hot}}$ (global IDs) \Comment{Phase 1}
\While{exp.\ pointer $\le$ position of $\mathrm{retset}[K'\!-\!1]$}
    \State expand first unvisited via $G_{\text{hot}}$; merge into $\mathrm{retset}$;
\EndWhile
\While{top-$K'$ change under $G_{\text{full}}$ expansion} \Comment{Phase 2}
    \State expand top-$K'$ of $\mathrm{retset}$ via $G_{\text{full}}$; merge into $\mathrm{retset}$;
\EndWhile
\State $\mathit{dist\_cnt} \gets 0$; $\mathit{next\_check} \gets \mathit{check\_period}$ \Comment{Phase 3}
\While{$\exists$ unvisited at position $k$ in $\mathrm{retset}$}
    \State $v \gets$ first unvisited
    \If{$v \in D_{\text{hot}}$ \textbf{and} $k < \tau_{\text{switch}}$}
        \State $\mathcal{N} \gets G_{\text{hot}}.\textsc{Neighbors}(v)$
    \Else
        \State $\mathcal{N} \gets G_{\text{full}}.\textsc{Neighbors}(v)$
    \EndIf
    \State merge $\mathcal{N}$ into $\mathrm{retset}$; trim to $L$
    \State $\mathit{dist\_cnt} \gets \mathit{dist\_cnt} + |\mathcal{N}|$
    \If{$\mathit{dist\_cnt} \ge \mathit{next\_check}$}
        \State assemble $\mathbf{f}$; \textbf{if} $T(\mathbf{f}) = 1$ \textbf{then break}
        \State $\mathit{next\_check} \gets \mathit{next\_check} + \mathit{check\_period}$
    \EndIf
\EndWhile
\State \Return top-$K$ of $\mathrm{retset}$
\end{algorithmic}
\end{algorithm}

\subsection{Complexity Analysis}
\label{sec:5-cplx}

We analyze the per-query search cost of the three-phase competitive search and the training cost of the decision tree.

\noindent\textbf{Search Cost.}
Let $IR = N_h / N$ denote the index ratio. A beam search on a single NSSG layer takes $O(\log n)$ expected distance computations on a graph of $n$ nodes~\cite{NSSG}. Each query first walks the hot graph in $O(\log(IR \cdot N))$ steps and, with probability $p_{\text{full}}$, continues into the full graph at cost $O(\log N)$:
\begin{equation}
C(IR) \;=\; \log(IR \cdot N) + p_{\text{full}} \cdot \log N,
\label{eq:cost}
\end{equation}
where $p_{\text{full}}$ is the probability that a query must reach $G_{\text{full}}$ during Phase~3. Under a Zipf workload with exponent $\alpha$, the frequency of rank-$i$ items is proportional to $i^{-\alpha}$. The cumulative mass on the top $N_h = IR \cdot N$ items is $\sum_{i=1}^{IR \cdot N} i^{-\alpha}/\sum_{i=1}^{N} i^{-\alpha}$. Approximating both sums by integrals,
\begin{equation}
p_{\text{full}} \;\approx\; 1 - \frac{1 - (IR \cdot N)^{1-\alpha}}{1 - N^{1-\alpha}}.
\label{eq:pfull}
\end{equation}
Substituting Eq.~\eqref{eq:pfull} into Eq.~\eqref{eq:cost} and differentiating with respect to $IR$,
\begin{equation}
\frac{dC}{dIR} \;=\; \frac{1}{IR} + \frac{(1-\alpha)\, N \log N \cdot (IR \cdot N)^{-\alpha}}{1 - N^{1-\alpha}};
\label{eq:dC}
\end{equation}
setting $dC/dIR = 0$ and isolating $IR$ yields the optimal ratio
\begin{equation}
IR^* \;=\; \left(\frac{N^{1-\alpha} - 1}{(1-\alpha)\, \log N \cdot N^{1-\alpha}}\right)^{\!1/(1-\alpha)}.
\label{eq:IRopt}
\end{equation}
For a representative workload ($N = 10^6$, $\alpha = 1.2$)~\cite{ZipfSearchEngine}, $IR^* \approx 0.002$, confirming that the optimal hot-set size is well below one percent. The closed form omits cache locality, decision-tree early stopping, and graph-degree effects, so $IR^*$ indicates the scale of the optimum rather than its exact value. The practical choice of $IR$ is dataset-dependent.

\noindent\textbf{Training Cost.}
Decision-tree fitting is $O(|S| \cdot 7 \cdot 10)$. At our defaults ($\mathit{check\_period} = 50$, $|Q_{\text{his}}| = 10^4$) $|S| \approx 10^5$--$10^6$ and training completes in seconds.

\section{Adaptive Hot Index Maintenance}
\label{sec:6-maint}
The dual-layer index and competitive search described in the preceding sections assume that the query distribution observed during construction remains representative at serving time. In practice, user interests and application workloads evolve continuously, causing a fixed $D_{\text{hot}}$ to gradually misalign with the actual query distribution. Periodically rebuilding the Hot Index from scratch would restore accuracy, incurring the full construction cost each time.

DQF avoids this cost by maintaining $G_{\text{hot}}$ incrementally. Each maintenance epoch involves four components, including a frequency tracker that identifies drifting nodes, a hit-rate detector that decides when to act, an incremental graph editor that applies structural changes, and decision-tree retraining.

\subsection{Frequency Tracking with Exponential Decay}
\label{sec:6-freq}

The frequency tracker specifies \emph{what to update} problem by selecting the updated nodes from  $D_{\text{hot}}$ as the workload evolves.
The constructed cumulative count $f(v)$ collected becomes outdated as query patterns shift. While sliding windows offer a basic solution, they treat all records equally and discard old data sharply. DQF instead maintains a \emph{decayed frequency score} $s_t(v)$ for each node, prioritizing recent activity and letting past data fade smoothly across boundaries.

Let $t$ denote the epoch number, $Q_t$ denote the queries served in epoch $t$, and $c_t(v)$ count $v$'s Top-$K$ appearances during $Q_t$. The decayed score updates as:
\begin{equation}
s_t(v) \;=\; \lambda^{\,\Delta t}\, s_{t_{\text{last}}(v)}(v) \;+\; c_t(v),
\label{eq:decay}
\end{equation}
with $\Delta t = t - t_{\text{last}}(v)$ and decay factor $\lambda \in (0,1)$ (default $\lambda = 0.5$). Rather than decaying every score each epoch at $O(N)$ cost, DQF applies \emph{lazy decay}: $s$ and $t_{\text{last}}$ are stored per node and updated only when $v$ is accessed, reducing the per-epoch cost to $O(|Q_t| \cdot K)$.

\subsection{Shift Detection via Hit-Rate Drop}
\label{sec:6-detect}

With decayed scores identifying which nodes have risen or fallen in popularity, the remaining question is \emph{when} to trigger a structural update. Triggering too frequently wastes compute on transient fluctuations, while triggering too late lets real drift degrade search quality unnoticed. An ideal signal should react to real workload changes but not to random fluctuations, require no ground-truth labels, and be cheap to compute online.

DQF monitors the \emph{hit rate} of the current Hot Index:
\begin{equation}
h_t \;=\; \frac{\sum_{q \in Q_t} |\mathrm{TopK}(q) \cap D_{\text{hot}}|}{|Q_t| \cdot K},
\label{eq:hitrate}
\end{equation}
which measures how often answers still reside in $D_{\text{hot}}$. The detector fires when $h_t$ drops relative to the recent baseline:
\begin{equation}
\text{trigger}_t \;=\; \mathbbm{1}\!\left[\,h_t \;<\; \gamma \cdot \bar{h}_{t-1}\,\right],
\label{eq:trigger}
\end{equation}
where $\bar{h}_{t-1}$ is the mean of the last $B_w = 2$ hit rates and $\gamma = 0.98$ by default.

\subsection{Incremental Hot-Index Update}
\label{sec:6-update}

Frequency tracking and hit-rate detection determine \emph{what} has drifted and \emph{when} to act. This subsection addresses the remaining question: \emph{how} to update $G_{\text{hot}}$ efficiently. The key challenge is to modify only the affected neighborhoods while preserving the angle-based diversity that NSSG provides.

When the detector triggers, DQF demotes fading nodes from $D_{\text{hot}}$, promotes rising ones, and repairs the affected edges. The procedure has three steps: \emph{asymmetric buffering} selects node candidates, \emph{hierarchical repair} fixes vacancies, and \emph{batch insertion} integrates the newcomers.

\subsubsection{Asymmetric Buffering}
\label{sec:6-asym}

Re-cutting $D_{\text{hot}}$ at rank $N_h$ each epoch would cause boundary nodes to be repeatedly inserted and removed. DQF introduces a buffer zone controlled by a lower ratio $\alpha_l$ and an upper ratio $\alpha_u$, where $\alpha_l = 0.75$ and $\alpha_u = 1.5$ by default:
\begin{itemize}[leftmargin=12pt,topsep=2pt,itemsep=1pt]
\item Insert $v \notin D_{\text{hot}}$ only if $\mathrm{rank}(v) \le \alpha_l \, N_h$.
\item Delete $v \in D_{\text{hot}}$ only if $\mathrm{rank}(v) > \alpha_u \, N_h$.
\item Nodes in $[\alpha_l \, N_h,\; \alpha_u \, N_h]$ are left in place.
\end{itemize}

\subsubsection{Hierarchical Graph Repair}
\label{sec:6-repair}

Removing a node $v_{\text{del}}$ from $G_{\text{hot}}$ is more difficult than insertion. Every in-neighbor of $v_{\text{del}}$ loses an out-edge, and if $v_{\text{del}}$ served as the sole bridge to a region of the graph, those neighbors also lose reachability to an entire cluster of nodes. Simply leaving the gap unfilled degrades both recall and routing efficiency. Re-pruning every affected node from scratch would restore full connectivity but costs nearly as much as a complete rebuild. In practice, ${\sim}80\%$ of affected nodes lose only one out-neighbor. For each affected in-neighbor $v$, let $\mathrm{lost}(v)$ denote the set of $v$'s out-neighbors that have been evicted from $D_{\text{hot}}$. DQF exploits this skew by splitting repair at a threshold $T_{\text{rep}} = 3$:

\begin{itemize}[leftmargin=12pt,topsep=2pt,itemsep=1pt]
\item \textit{Lightweight repair} ($|\mathrm{lost}(v)| < T_{\text{rep}}$): the surviving adjacency list of $v$ is \emph{kept in place}. Only the deleted nodes' original neighbors enter the candidate pool and are sorted by L2 distance to $v$. Candidates that pass angle pruning against the surviving neighbors fill the empty slots.
\item \textit{Full repair} ($|\mathrm{lost}(v)| \ge T_{\text{rep}}$): the surviving neighbors \emph{and} the deleted nodes' neighbors are merged into a single candidate pool. The pool is sorted and NSSG-pruned from scratch, \emph{replacing} $v$'s adjacency list entirely. This is equivalent to re-running local NSSG construction for $v$.
\end{itemize}

Both paths apply the same angle threshold $\theta$ used during initial construction, ensuring that repaired edges remain compatible with the original NSSG topology.

\subsubsection{Batch Insertion}
\label{sec:6-insert}

After repair, each newly admitted node $v_{\text{new}} \in \mathcal{I}$ is inserted into $G_{\text{hot}}$ via beam search on the partially-repaired graph. Angle-based pruning selects its out-neighbors, and reverse edges are written back, dropping the farthest edge when the degree exceeds $R$.

\subsection{Decision Tree Retraining}
\label{sec:6-retrain}

The decision tree learns its stopping thresholds from the feature distributions observed under the training query workload. Once the query distribution shifts, those feature distributions change as well, causing the tree to stop search too early or too late. Retraining is controlled by the same hit-rate trigger as the graph update (Eq.~\eqref{eq:trigger}). When it fires, DQF samples $N_{\text{train}} = 5\,000$ queries from the current epoch's log and refits the tree using the same pipeline.

\begin{algorithm}[tbp]
\caption{Epoch Maintenance}
\label{alg:maint}
\begin{algorithmic}[1]
\Require epoch index $t$, queries $Q_t$, dual-layer index $(G_{\text{full}}, G_{\text{hot}})$, tracker state $(s, t_{\text{last}})$, detector history $\bar{h}$, tree $T$.
\Ensure updated $(G_{\text{hot}}, T)$ and refreshed tracker/detector state.
\State for each $v$ hit during $Q_t$: refresh $s_t(v)$ via Eq.~\eqref{eq:decay} \Comment{lazy decay}
\State $h_t \gets$ hit rate of $Q_t$ on $D_{\text{hot}}$ \Comment{Eq.~\eqref{eq:hitrate}}
\If{$h_t \ge \gamma \cdot \bar{h}_{t-1}$}
    \State \Return \Comment{Eq.~\eqref{eq:trigger}}
\EndIf
\State rank all nodes by $s_t$ (descending)
\State $\mathcal{D} \gets \{v \in D_{\text{hot}} : \mathrm{rank}(v) > 1.5 N_h\}$ \Comment{evictions}
\State $\mathcal{I} \gets \{v \notin D_{\text{hot}} : \mathrm{rank}(v) \le 0.75 N_h\}$ \Comment{promotions}
\For{each $v_{\text{del}} \in \mathcal{D}$}
    \State remove $v_{\text{del}}$ and record $\mathrm{lost}(\cdot)$ on its in-neighbors
\EndFor
\For{each affected $v$}
    \If{$|\mathrm{lost}(v)| < T_{\text{rep}}$}
        \State \Call{LightweightRepair}{$v$}
    \Else
        \State \Call{FullRepair}{$v$}
    \EndIf
\EndFor
\For{each $v_{\text{new}} \in \mathcal{I}$}
    \State $C \gets$ \Call{BeamSearch}{$G_{\text{hot}}, v_{\text{new}}, L$}
    \State $N(v_{\text{new}}) \gets$ \Call{AnglePrune}{$C$}
    \State write reverse edges; trim each neighbor's list to $R$
\EndFor
\State repack $G_{\text{hot}}$ adjacency into optimized binary
\State $T \gets$ \Call{RetrainTree}{sample of $N_{\text{train}}$ from $Q_t$}
\State update $\bar{h} \gets h_t$
\end{algorithmic}
\end{algorithm}

Algorithm~\ref{alg:maint} summarizes the full epoch maintenance procedure. The first guard (lines~3--4) checks whether the hit rate remains above the threshold. When the guard fires, lines~5--7 rank every node by the decayed frequency score and identify the eviction set $\mathcal{D}$ and the promotion set $\mathcal{I}$. Lines~8--14 remove the evicted nodes and repair their former in-neighbors. Lines~15--18 insert each promoted node via beam search on the partially repaired graph, followed by angle pruning and reverse-edge write-back. Finally, lines~19--21 repack the updated adjacency lists and retrain the decision tree on a fresh query sample from the current epoch.

\subsection{Complexity Analysis}
\label{sec:6-cplx}

We analyze the per-epoch cost of the maintenance pipeline to show that incremental updates are far cheaper than full reconstruction.

The monitoring overhead (frequency tracking and hit-rate detection) is $O(|Q_t| \cdot K)$, collected during search with no extra I/O. When the detector fires, the structural update cost is dominated by repair and insertion:
\begin{equation}
O\!\left(|\mathcal{D}| \cdot \bar{d}^2 \;+\; |\mathcal{I}| \cdot L\right),
\label{eq:upd-cost}
\end{equation}
where each evicted node affects up to $\bar{d}$ in-neighbors and patching each costs $O(\bar{d})$. Decision-tree retraining adds $O(N_{\text{train}} \cdot L_{\text{train}})$ for replaying $N_{\text{train}} = 5\,000$ queries. The total per-maintenance cost is
\begin{equation}
\underbrace{O\!\left(|\mathcal{D}| \cdot \bar{d}^2 + |\mathcal{I}| \cdot L\right)}_{\text{structural repair}} \;+\; \underbrace{O\!\left(N_{\text{train}} \cdot L_{\text{train}}\right)}_{\text{tree replay {\&} fit}}.
\label{eq:total-maint}
\end{equation}
The key observation is that both terms scale with the number of \emph{changed} nodes ($|\mathcal{D}|$, $|\mathcal{I}|$), not the full index size $N_h$. A full rebuild, by contrast, must re-run KNNG construction and NSSG pruning at $O(N_h^{1.16} + d\,N_h\,r\,k^2)$ and replay the entire query log. Under stable drift, the asymmetric buffer keeps $|\mathcal{D}|, |\mathcal{I}| \ll N_h$, so each incremental update costs only a small fraction of a rebuild.

\begin{table}[tbp]
\centering
\vspace{-0.1in}
\caption{Dataset Statistics.}
\vspace{-0.05in}
\label{tab:dataset}
\resizebox{0.49\textwidth}{!}{
\begin{tabular}{@{}lccccc@{}}
\toprule
\textbf{Dataset} & \textbf{Dim.} & \textbf{Base Size} & \textbf{Database} & \textbf{Query Pool} & \textbf{Data Type} \\
\midrule
SIFT1M   & 128  & 1,000,000  & 900,000   & 100,000   & Image descriptors \\
GIST1M   & 960  & 1,000,000  & 900,000   & 100,000   & Image descriptors \\
Crawl    & 300  & 1,989,995  & 1,790,996 & 198,999   & Web embeddings \\
GloVe    & 100  & 1,183,514  & 1,065,163 & 118,351   & Word embeddings \\
Deep100M & 96   & 100,000,000 & 90,000,000 & 10,000,000 & Image embeddings \\
\bottomrule
\end{tabular}
}
\vspace{-0.1in}
\end{table}

\section{Experimental Evaluation}
\label{sec:7-exp}

We evaluate DQF along four axes, including construction overhead, end-to-end search performance, robustness under distribution drift, and sensitivity to design choices and workload parameters. All experiments use the setup described below unless stated otherwise.

\begin{table*}[t]
\centering
\belowrulesep=0pt
\aboverulesep=0pt
\caption{Index Construction Time (seconds) and Index Size (MB). \\DQF overhead represents the additional cost on top of NSSG (total DQF cost = NSSG + DQF overhead).}
\vspace{-0.1in}
\label{tab:build_storage}
\resizebox{\textwidth}{!}{
\begin{tabular}{@{}l|cccccccc|cccccccc@{}}
\toprule
& \multicolumn{8}{c|}{\textbf{Build Time (seconds)}} & \multicolumn{8}{c}{\textbf{Index Size (MB)}} \\
\textbf{Dataset} & Annoy & HNSW & NSSG & LSHAPG & FLANN & NSG & Vamana & \textbf{DQF} & Annoy & HNSW & NSSG & LSHAPG & FLANN & NSG & Vamana & \textbf{DQF} \\
\midrule
SIFT1M  & 54.0  & 115.8  & 80.8   & 85.6   & 161.1  & 119.3  & 38.80  & \textbf{80.8 + 2.61}  & 4020.0 & 189.2 & 137.4 & 409.9  & 211.7 & 102.8 & 139.37 & \textbf{137.4 + 2.20} \\
GIST1M  & 283.0 & 1109.4 & 695.6  & 201.5  & 1984.0 & 1254.8 & 645.77 & \textbf{695.6 + 20.64} & 4032.8 & 257.8 & 123.0 & 381.9  & 960.5 & 76.9  & 157.88 & \textbf{123.0 + 16.48} \\
Crawl   & 260.6 & 845.2  & 1029.0 & 342.7  & 1606.2 & 2252.8 & 256.78 & \textbf{1029.0 + 14.01} & 8060.1 & 335.5 & 160.3 & 740.0  & 902.4 & 86.5  & 187.17 & \textbf{160.3 + 10.38} \\
GloVe   & 71.7  & 705.5  & 693.7  & 156.9  & 2661.6 & 785.7  & 64.30  & \textbf{693.7 + 12.03} & 6038.4 & 240.1 & 92.0  & 428.9  & 141.1 & 57.1  & 105.51 & \textbf{92.0 + 2.11} \\
\bottomrule
\end{tabular}
}
\vspace{-0.05in}
\end{table*}

\begin{figure*}[t]
    \centering
    \includegraphics[width=\textwidth]{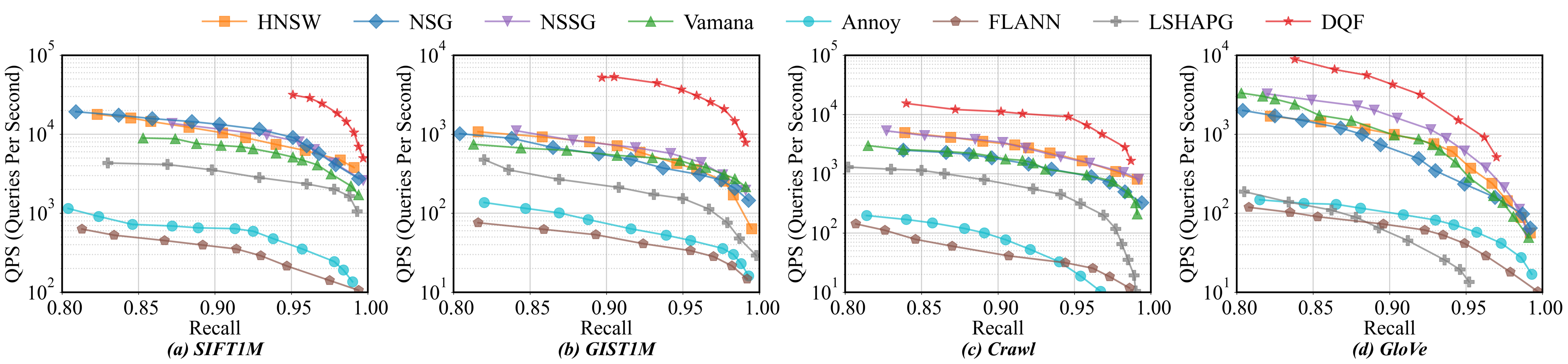}
    \vspace{-0.8cm}
    \caption{QPS--Recall@10 Pareto frontiers on the four million-scale datasets (higher and farther right is better).}
    \label{fig:qps_recall}
    \vspace{-0.5cm}
\end{figure*}

\begin{figure*}[t]
    \centering
    \includegraphics[width=\textwidth]{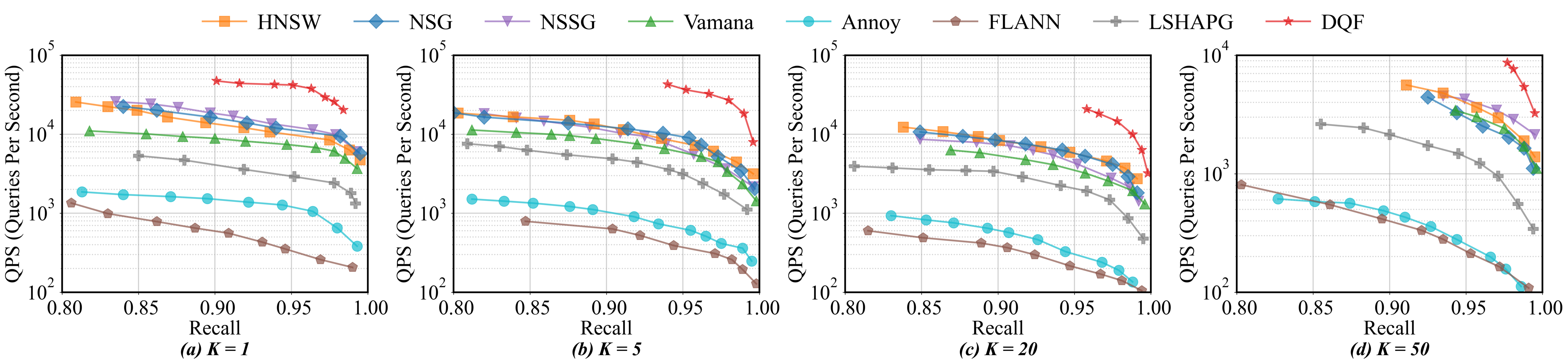}
    \vspace{-0.6cm}
    \caption{Impact of Result Size $K$ on SIFT1M}
    \label{fig:k_sensitivity}
    \vspace{-0.5cm}
\end{figure*}

\subsection{Experimental Setup}
\label{sec:exp_setup}

\textbf{Datasets.} Table~\ref{tab:dataset} summarizes the five datasets, spanning 96--960 dimensions and 1M--100M vectors. Each is split at the ratio of 9/1 into database $\mathcal{D}$ and query pool $\mathcal{Q}$. To model real-world frequency skew, query indices are sampled from $\mathcal{Q}$ following a Zipf distribution ($\alpha = 1.2$)~\cite{ZipfSearchEngine}. Each sampled query is then perturbed with 5\% Gaussian noise, because semantically equivalent queries often map to different but nearby embedding vectors (Section~\ref{sec:3-back}). We sample from the held-out pool $\mathcal{Q}$ rather than from the indexed database $\mathcal{D}$, ensuring that no test query appears in the index~\cite{wang2024starling}. Popular database vectors are still retrieved frequently because semantically close queries tend to hit the same targets. To our knowledge, no public ANNS benchmark provides per-query frequency labels. The Zipf-plus-noise scheme above is therefore the most practical way to vary both query skew and drift in a controlled setting. We vary $\alpha$ and noise in Section~\ref{sec:exp_robustness}. Each experiment uses 1,000 test queries.

\textbf{Baselines.} We compare against seven SOTA methods spanning three paradigms. \textit{Graph-based:} HNSW~\cite{HNSW} (the widely adopted industry standard), NSG~\cite{NSG} (MRNG-pruned sparse graph), NSSG~\cite{NSSG} (angle-based NSSG pruning), and Vamana~\cite{Diskann} (DiskANN's in-memory graph with $\alpha$-RobustPrune). \textit{Tree-based:} Annoy~\cite{Annoy} (random projection forest) and FLANN~\cite{FLANN} (randomized KD-tree ensemble). \textit{Hybrid:} LSH-APG~\cite{LSHAPG} (locality-sensitive hashing with proximity graph). All methods use identical hardware, data splits, and queries. Search is single-threaded, and each method's parameters are grid-tuned to produce Pareto-optimal QPS--Recall curves. Note that PANNS~\cite{PANNS} and DeDrift~\cite{Dedrift} discussed in Section~\ref{sec:2-rel-dynamic} are not included because they assume hotness is driven by recent insertion rather than content-driven query preferences, and neither has released source code, which makes a fair comparison infeasible.

\textbf{Implementation Details.}
Source code is publicly available on GitHub.\footnote{\url{https://github.com/ZJU-DAILY/DQF}}
All methods are compiled with GCC (\texttt{-O3 -march=native}), use auto-vectorized SIMD, and run on an Intel Xeon Silver 4310 (2.10\,GHz, 1\,TB RAM). Index construction uses 48-thread OpenMP; all query measurements are single-threaded and memory-resident.

\begin{figure*}[t]
    \centering
    \includegraphics[width=0.99\textwidth]{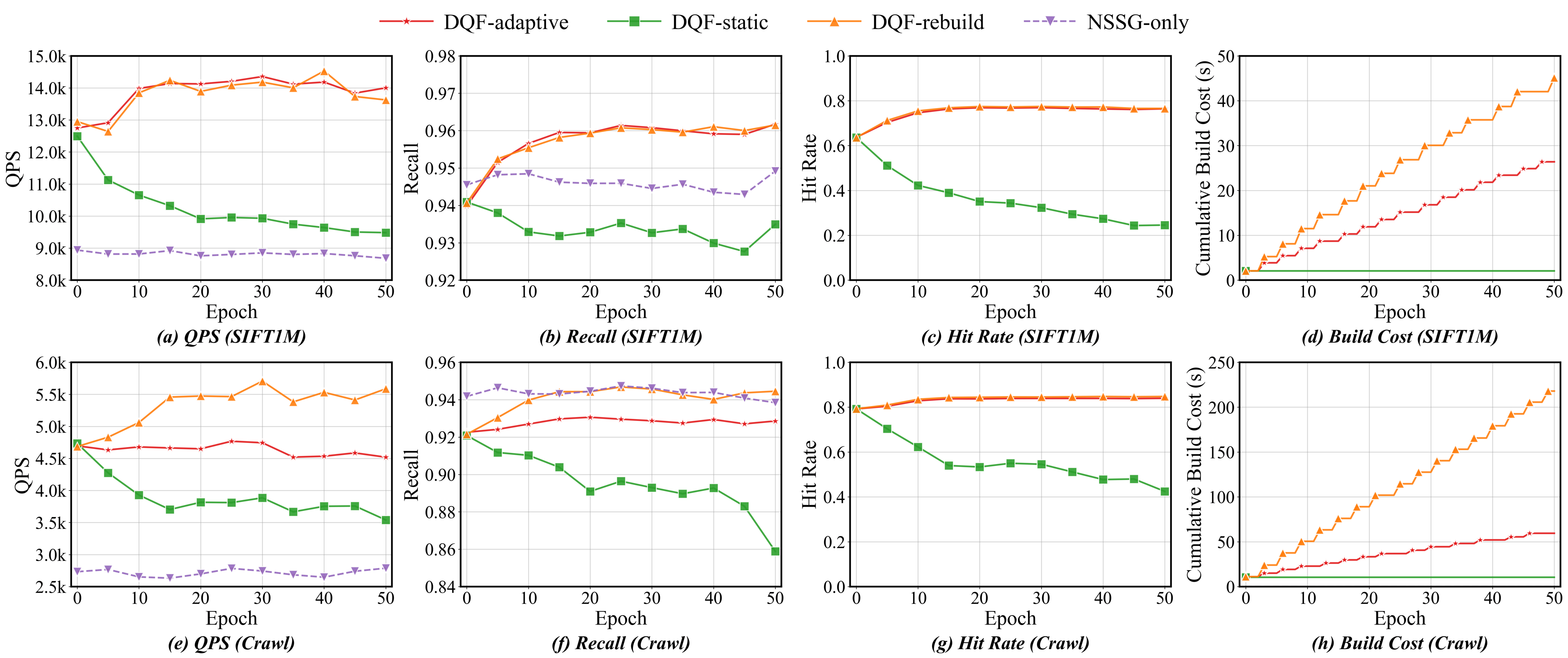}
    \vspace{-0.3cm}
    \caption{Adaptive maintenance evaluation on SIFT1M (top) and Crawl (bottom) over 50 epochs of progressive query distribution shift.}
    \label{fig:shift}
    \vspace{-0.5cm}
\end{figure*}

\textbf{Evaluation Metrics.} We adopt Recall@$K$ (fraction of true $K$-nearest neighbors returned) to measure the result quality, and QPS (queries per second) to measure the throughput.

\subsection{Index Construction and Storage Overhead}
\label{sec:exp_construction}

A query-aware index is only practical if it does not introduce excessive construction or storage costs. Table~\ref{tab:build_storage} shows that on the four million-scale datasets, DQF's overhead on top of NSSG is marginal: 1.4--3.2\% additional build time and 1.6--13.4\% additional index size (Hot Index + decision tree).

\begin{figure*}[t]
    \centering
    \includegraphics[width=\textwidth]{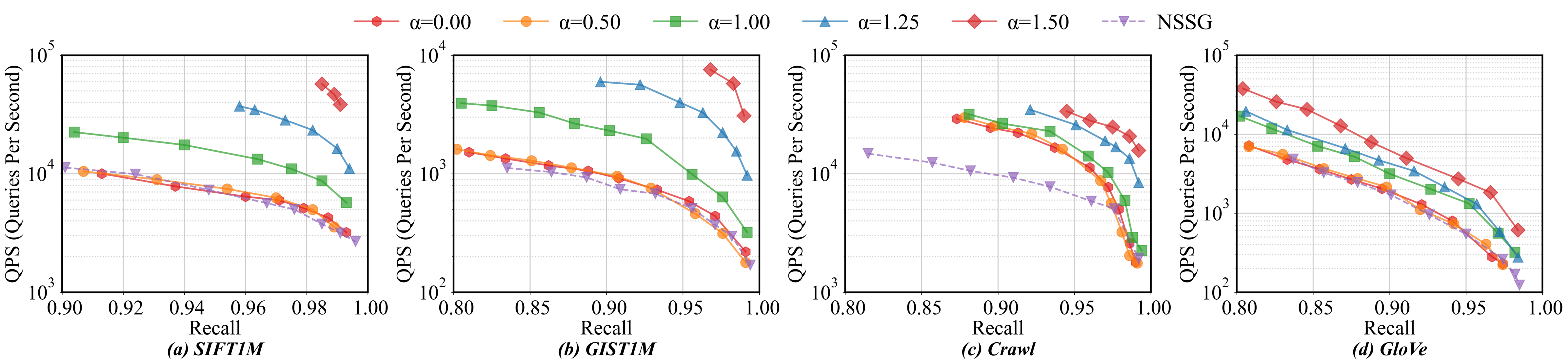}
    \vspace{-0.8cm}
    \caption{Speedup vs.\ Workload Skewness $\alpha$}
    \label{fig:alpha_sensitivity}
    \vspace{-0.3cm}
\end{figure*}

\subsection{Overall Performance Comparison}
\label{sec:exp_overall}

Figure~\ref{fig:qps_recall} plots QPS against Recall@10 for all methods on the four million-scale datasets. On all four datasets, DQF dominates the Pareto frontier. At 95\% recall, DQF delivers a 2.2--6.9$\times$ speedup over the strongest baseline on each dataset: 3.4$\times$ over NSG on SIFT1M, 6.9$\times$ over NSSG on GIST1M, 4.6$\times$ over HNSW on Crawl, and 2.2$\times$ over NSSG on GloVe. On two of the four datasets, NSSG (DQF's own base graph) is already the strongest baseline. The speedup therefore directly quantifies the gain from query-aware optimization. The largest gain appears on GIST1M (960D), where the compact Hot Index benefits most from cache locality. The smallest gain is on GloVe. GloVe vectors have only 100 coordinates, but the data spans many independent directions (high intrinsic dimensionality). As a result, neighbors are scattered across more regions, and the Hot Index can cover a smaller fraction of them. These two datasets show when DQF helps most and least: the benefit grows when frequent queries cluster in a compact region of the vector space.

Among baselines, graph-based methods (HNSW, NSG, Vamana) consistently outperform tree-based (Annoy, FLANN) and hybrid (LSH-APG) approaches at high recall, confirming prior findings~\cite{wang2024starling}. DQF adopts NSSG as its base graph, which is already among the strongest single graphs. The query-aware layer further boosts its throughput, allowing DQF to outperform every competing method on all four datasets. This demonstrates that the gains come from query-aware optimization rather than from the underlying graph topology.

\subsection{Adaptive Hot Index Maintenance}
\label{sec:exp_shift}

A static Hot Index inevitably becomes outdated as the query distribution drifts. To simulate this, we perturb the popularity ranking of query-pool nodes each epoch using a Gaussian random walk. Queries are then re-drawn from the updated ranking via the Zipf generator. Because Zipf $\alpha{=}1.2$ assigns most query traffic to the top-ranked items, even moderate rank perturbations quickly change which objects are hot. We evaluate over 50 epochs on SIFT1M (128D) and Crawl (300D), issuing $0.01|\mathcal{D}|$ queries per epoch. Before drift begins, DQF is initialized with five warm-up epochs of NSSG queries to obtain the initial frequency distribution and build the Hot Index. Four strategies are compared: \textbf{DQF-adaptive} (incremental updates triggered by hit-rate drop), \textbf{DQF-static} (built once, never maintained), \textbf{DQF-rebuild} (full Hot Index reconstruction on each detected drift), and \textbf{NSSG-only} (no Hot Index). Decision-tree retraining cost is included in Figure~5(d). The central question is whether lightweight incremental maintenance can match full reconstruction in search quality while costing significantly less. Figure~\ref{fig:shift} presents the results.

\textbf{Throughput and Recall.} DQF-adaptive and DQF-rebuild both sustain stable throughput throughout 50 epochs, consistently reaching 1.6--1.7$\times$ the speed of NSSG-only on both datasets. DQF-static, in contrast, starts at the same level but steadily degrades as drift accumulates, eventually approaching the NSSG-only baseline. The recall curves reveal a further consequence: while the maintained strategies all stay above 0.92, DQF-static's recall visibly drops on Crawl, because its outdated decision tree makes increasingly inaccurate early-stopping decisions. In other words, failing to maintain the Hot Index hurts not only throughput but also result quality.

\begin{figure*}[t]
    \centering
    \vspace{-0.4cm}
    \includegraphics[width=\textwidth]{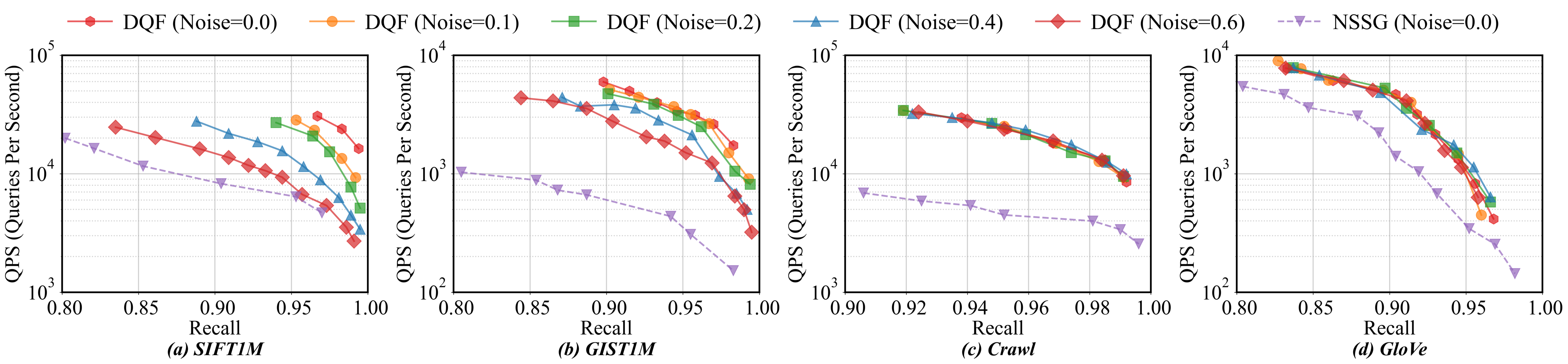}
    \vspace{-0.8cm}
    \caption{QPS and Recall vs.\ Query Perturbation Noise}
    \vspace{-0.3cm}
    \label{fig:noise_sensitivity}
\end{figure*}

\begin{figure*}[t]
    \centering
    \includegraphics[width=\textwidth]{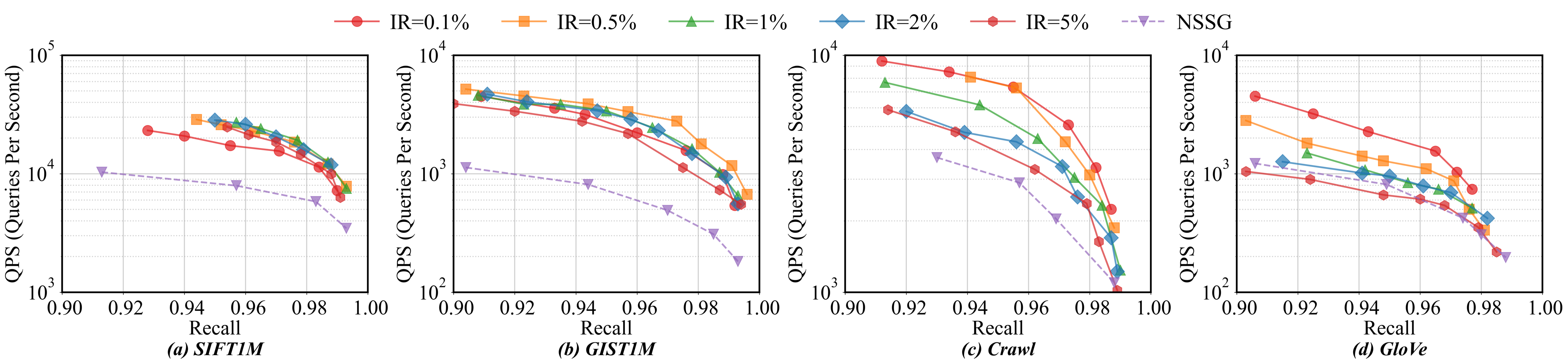}
    \vspace{-0.8cm}
    \caption{QPS-Recall@10 under varying Index Ratio $IR \in \{0.1\%, 0.5\%, 1\%, 2\%, 5\%\}$}
    \label{fig:ir_sensitivity}
    \vspace{-0.6cm}
\end{figure*}

\textbf{Hit Rate Decay.} Figure~5(c) explains the degradation above. Without maintenance, the Hot Index gradually loses relevance to the current query distribution: DQF-static's hit rate drops by more than half on both datasets, meaning most queries bypass the compact Hot Index entirely and must be served by the slower Full Index. DQF-adaptive avoids this by incrementally replacing outdated entries, keeping the hit rate stable throughout the experiment.

\textbf{Maintenance Cost.} Since DQF-adaptive and DQF-rebuild deliver comparable search quality, the deciding factor is maintenance cost. DQF-adaptive is significantly cheaper, reducing cumulative build time by \textbf{45\%} on SIFT1M and \textbf{76\%} on Crawl relative to full reconstruction. The gap widens on larger datasets because incremental repair touches only the changed portion of the Hot Index, while full reconstruction must rebuild the entire subgraph.

These results confirm that adaptive maintenance is the practical choice: it matches full reconstruction in search quality while only requiring cheap, local updates.

\begin{figure*}[t]
    \centering
    \includegraphics[width=\textwidth]{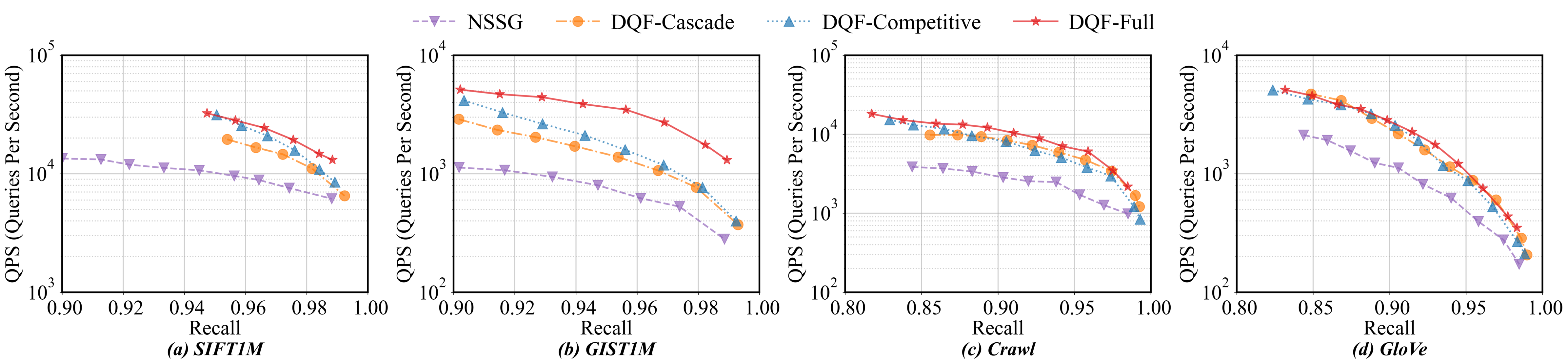}
    \vspace{-0.8cm}
    \caption{Ablation study: QPS-Recall@10 curves comparing NSSG (base), DQF-Cascade (dual-index with sequential search), DQF-Competitive (three-phase competitive search, no decision tree), and DQF-Full (complete framework).}
    \vspace{-0.4cm}
    \label{fig:ablation}
\end{figure*}

\begin{figure}[t]
    \centering
    \vspace{-0.3cm}
    \includegraphics[width=0.44\textwidth]{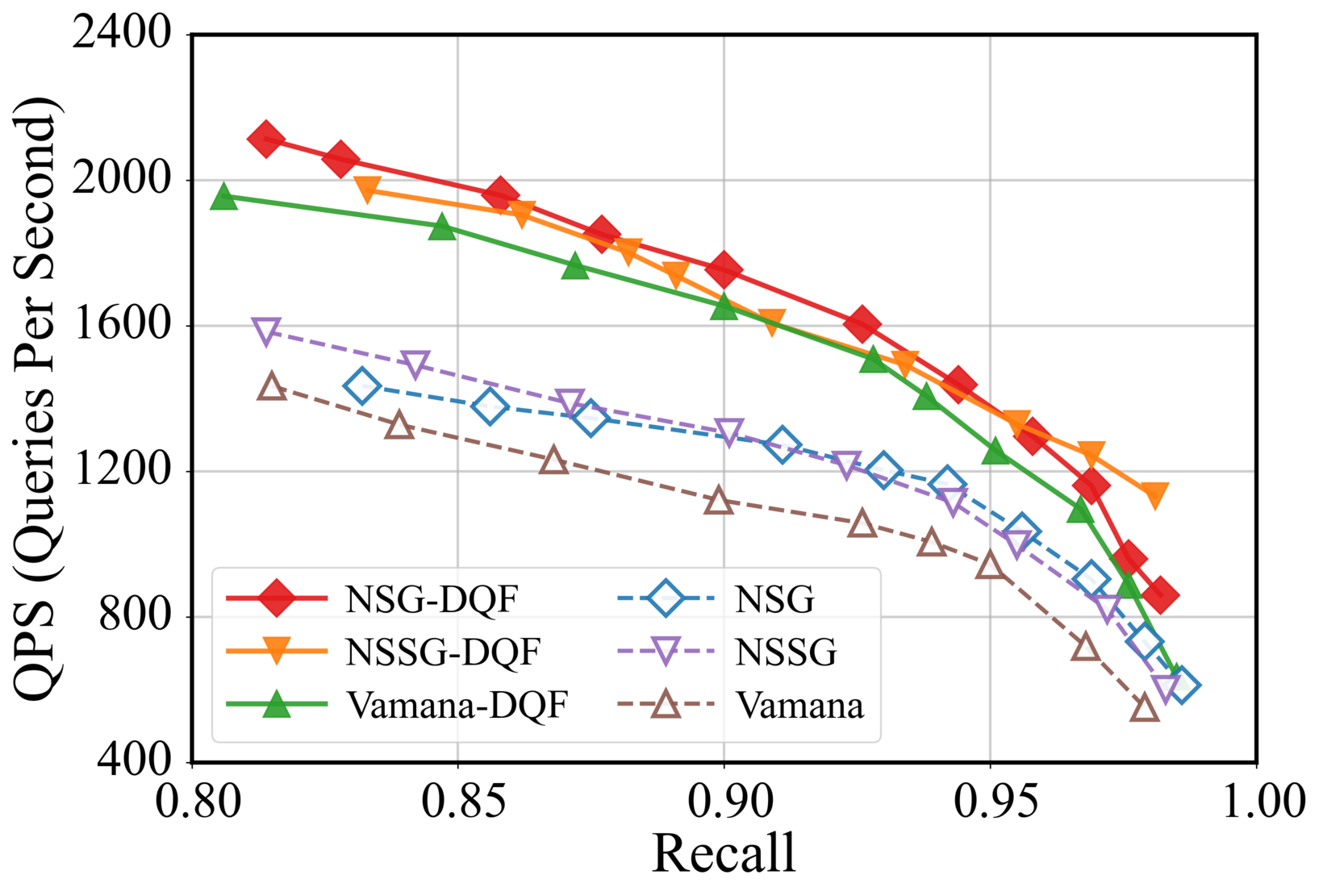}
    \vspace{-0.4cm}
    \caption{QPS-Recall Curve on Deep100M}
    \vspace{-0.6cm}
    \label{fig:deep}
\end{figure}

\subsection{Robustness Analysis}
\label{sec:exp_robustness}

We assess DQF's robustness by varying four key workload and configuration parameters: result size $K$, query skewness $\alpha$, query perturbation noise, and index ratio $IR$.

\textbf{Sensitivity to Result Size $K$.} Figure~\ref{fig:k_sensitivity} sweeps $K$ on SIFT1M. DQF consistently outperforms NSSG at all four settings, with the largest gains at moderate $K$ ($K{=}5$ and $K{=}20$). When $K$ is small, the base search itself is fast, leaving limited room for acceleration. When $K$ is large, the Hot Index ($IR{=}0.5\%$, roughly 4{,}500 vectors) can cover only a small fraction of the $K$ true neighbors, so most work falls back to the Full Index and the benefit of the compact subgraph shrinks. At $K{=}50$, even with the smallest feasible search budget ($L{=}K{=}50$), DQF exceeds 97\% recall, showing that DQF remains effective.

\textbf{Sensitivity to Query Skewness $\alpha$.} Figure~\ref{fig:alpha_sensitivity} varies the Zipf exponent $\alpha \in \{0.0, 0.5, 1.0, 1.25, 1.5\}$, which controls how concentrated query mass is on popular items. DQF's speedup grows super-linearly with $\alpha$: at $\alpha{=}0$ (uniform workload) DQF achieves nearly the same throughput as NSSG, confirming that the additional components introduce no overhead when there is no skew to exploit. In the production-realistic range ($\alpha \in [1.0, 1.5]$)~\cite{ZipfSearchEngine}, with the decision tree retrained per $\alpha$, it delivers substantial gains.

\textbf{Sensitivity to Query Perturbation.}
Figure~\ref{fig:noise_sensitivity} confirms the stability observation of Section~\ref{sec:3-back}: acceleration should come from hot \emph{objects}, not from caching exact query vectors. DQF consistently outperforms NSSG across all noise levels. The speedup decreases as the perturbation increases, but remains positive even at extreme noise. This is because the Hot Index caches the target region of embedding space rather than exact query vectors, and the Full Index always guarantees coverage.

\textbf{Sensitivity to Index Ratio $IR$.} The index ratio $IR = |\mathcal{D}_h|/|\mathcal{D}|$ controls the Hot Index size and governs the trade-off between subgraph coverage and search efficiency. Figure~\ref{fig:ir_sensitivity} sweeps $IR$ across all four datasets. The best $IR$ varies by dataset: SIFT and GIST peak around $IR \in [0.5\%, 1\%]$, while Crawl and GloVe prefer smaller values where the compact Hot Index fits better in cache. On all datasets, $IR$ beyond $2\%$ hurts throughput because the Hot Index becomes too large to retain its locality benefit. We use $IR = 0.5\%$ as the default because it gives consistent speedup on every dataset with only 2--17\,MB extra storage, though tuning $IR$ per dataset can bring additional gains.

\subsection{Ablation Study}
\label{sec:exp_ablation}

To quantify each component's contribution, we evaluate four configurations (Figure~\ref{fig:ablation}): \textbf{NSSG} (base graph only), \textbf{DQF-Cascade} (dual-index with sequential hot-then-full search), \textbf{DQF-Competitive} (three-phase competitive search without decision tree), and \textbf{DQF-Full} (complete framework).

\textbf{Dual-Index (NSSG $\to$ Cascade).} Adding the Hot Index alone yields 1.9--2.8$\times$ speedup at 95\% recall. The gain comes purely from caching frequently accessed vectors in a compact subgraph, validating the dual-layer design of Section~\ref{sec:4-index}.

\textbf{Competitive Search (Cascade $\to$ Competitive).} Source-aware expansion (Section~\ref{sec:competitive-search}) adds a further 1.2--1.6$\times$ on SIFT and GIST, where candidates from the Hot Index and Full Index cover different parts of the neighbor set. 
\textbf{Decision Tree (Competitive $\to$ Full).} Learned early stopping provides the largest incremental gain on high-dimensional datasets (2.1$\times$ on GIST, 1.5$\times$ on Crawl). On these datasets, the gain more than offsets the extra cost of competitive search.

\textbf{Feature Importance.} A decision tree trained at ${\sim}$95\% recall (Table~\ref{tab:feature_importance}) shows dynamic features ($f_3$--$f_7$) account for 75--78\% of total importance, confirming that dynamic features computed during search are more predictive than static features estimated before search ($f_1, f_2$). The top three are current-result quality ($f_3$), hot-coverage ratio ($f_7$), and cumulative distance count ($f_5$), though relative rankings shift across datasets (e.g., $f_7$ dominates on SIFT and GIST, while GloVe favors $f_4$--$f_5$). Every feature exceeds 7\% on at least one dataset, justifying the full seven-feature set.

\textbf{Effect of Three-tier Sample Weighting.} The three-tier sample weighting (Section~\ref{sec:dt-weight}) does not target throughput. Its role is to preserve recall in the high-recall regime. Without it, training samples labeled \emph{continue searching} become very rare in the high-recall region, so a uniformly-weighted tree learns to stop too aggressively and recall saturates near 95\%, unable to reach the 98--99\% range.

\textbf{Effect of Frequency Weighting in Graph Construction.} In a supplementary ablation, disabling the frequency weighting ($w(v){=}1$ in Eq.~\ref{eq:cmp}) consistently lowers QPS by ${\sim}$10\% on GloVe and 3--5\% on the remaining datasets, with no change in recall. The weighting only modifies graph construction, making the graph structure more favorable for frequently queried regions. Since it adds nothing at query time, the throughput gain comes at zero additional cost.

In summary, the Hot Index delivers high-quality seed candidates, the competitive search exploits dual-layer synergy, and the decision tree adapts search depth per query.

\subsection{Performance on Large-Scale Data}
\label{sec:exp_scale}

To test whether DQF's gains transfer beyond million-scale, we couple it with three base graphs (NSSG, NSG, Vamana) on Deep100M (100M vectors, 96D). Figure~\ref{fig:deep} shows all DQF variants dominate their respective base graphs: ${\sim}$1.3$\times$ at 95\% recall, widening to 1.5--1.7$\times$ at ${\geq}$98\% recall, where search takes more iterations to converge, and the decision tree's early termination saves more work. The smaller relative speedup compared to million-scale is expected: at 100M the graph is much larger, so the search requires more hops to reach the target region before it begins refining candidates. DQF's Hot Index and early stopping mainly accelerate the refinement stage, which now accounts for a smaller share of total cost. Since DQF is built on top of flat base graphs, the fair comparison is between each base graph and its DQF-augmented variant. All three flat graphs benefit, confirming that DQF's query-aware optimization is independent of the base graph's pruning strategy. HNSW is not included as a base graph. DQF's competitive search is designed for single-layer flat graphs and cannot directly merge candidates across HNSW's multiple routing layers. Adapting the algorithm to hierarchical graphs is left as future work.

\begin{table}[t]
\centering
\vspace{-0.1in}
\caption{Decision-tree feature importance (\%, Gini) at ${\sim}$95\% baseline recall.  Features sorted by cross-dataset average.}
\label{tab:feature_importance}
\vspace{-0.1in}
\begin{tabular}{lccccc}
\toprule
\textbf{Feature} & \textbf{SIFT} & \textbf{GIST} & \textbf{GloVe} & \textbf{Crawl} & \textbf{Avg} \\
\midrule
$f_3$: \texttt{dist\_1st}        & 18.6 & 19.8 & 23.5 & \textbf{24.1} & 21.5 \\
$f_7$: \texttt{hot\_ratio\_topk} & \textbf{27.3} & \textbf{24.9} & 8.0  & 13.2 & 18.3 \\
$f_5$: \texttt{dist\_cnt}        & 14.8 & 13.1 & \textbf{24.4} & 18.3 & 17.7 \\
$f_1$: \texttt{hot\_dist\_1st}   & 16.2 & 17.4 & 13.8 & 14.1 & 15.4 \\
$f_4$: \texttt{dist\_ratio\_1\_k}& 6.9  & 7.5  & 14.7 & 12.0 & 10.3 \\
$f_6$: \texttt{update\_cnt}      & 8.7  & 10.2 & 7.3  & 10.0 & 9.1  \\
$f_2$: \texttt{hot\_ratio\_1\_k} & 7.6  & 7.3  & 8.3  & 8.3  & 7.9  \\
\bottomrule
\end{tabular}
\vspace{-0.12in}
\end{table}

\section{Conclusion}
\label{sec:8-concl}
We presented DQF, a Dual-Index Query Framework that makes graph-based ANNS workload-aware. A frequency-weighted Hot Index, three-phase competitive search with decision-tree early stopping, and hit-rate-triggered incremental maintenance deliver 2.2--6.9$\times$ speedups over the strongest baseline at the million scale (95\% recall) and scale to 100M vectors with consistent gains. They also sustain the advantage under query drift at a fraction of the rebuild cost and improve three flat-graph base indexes. Future directions include porting to hierarchical indexes, validation on industrial query logs whose richer temporal structure goes beyond the stationary Zipf model used here.

\bibliographystyle{IEEEtran}
\bibliography{sample}

\end{document}